\documentclass[
    ,final            
  ]
  {aipproc}

\layoutstyle{6x9}

\def\ca{\c{c}\~{a}}

\usepackage[ansinew]{inputenc}
\usepackage{amsfonts}
\usepackage{epsfig}
\usepackage{color}

\def\mathbf{\vec}
\def\be{\begin{equation}}
\def\ee{\end{equation}}
\def\ba{\begin{eqnarray}}
\def\ea{\end{eqnarray}}

\def\bx
  {\hbox to.77778em{\hfil\vrule\vbox to.675em
  {\hrule width.6em\vfil\hrule}\vrule\hfil}}

\begin{document}

\title{Stable Multiquark Interactions}


\author{B. Hiller}{
  address={Centro de F\'{\i}sica Te\'{o}rica, Departamento de
         F\'{\i}sica da Universidade de Coimbra, 3004-516 Coimbra, Portugal}
}

\author{A.A. Osipov}{
  address={Centro de F\'{\i}sica Te\'{o}rica, Departamento de
         F\'{\i}sica da Universidade de Coimbra, 3004-516 Coimbra, Portugal}
 ,altaddress={Joint Institute of Nuclear Research, Laboratory of Nuclear Problems, 141980 Dubna, Moscow Region, Russia}
}
\author{A.H. Blin}{
  address={Centro de F\'{\i}sica Te\'{o}rica, Departamento de
         F\'{\i}sica da Universidade de Coimbra, 3004-516 Coimbra, Portugal}
 }
\author{J. da Provid\^encia}{
  address={Centro de F\'{\i}sica Te\'{o}rica, Departamento de
         F\'{\i}sica da Universidade de Coimbra, 3004-516 Coimbra, Portugal}
}

\begin{abstract}
The necessity of adding higher order multiquark interactions to the three flavor NJL model with $U_A(1)$ breaking,   in order to stabilize its vacuum, is discussed.
\end{abstract}

\maketitle



It has recently been shown \cite{Osipov1:2006} that the $SU(3)_L \times SU(3)_R$ chiral symmetric Lagrangian with axial $U_A(1)$ breaking composed of the $4q$ ($q=$quark) Nambu-Jona-Lasinio (NJL) \cite{NJL:1961} and $2N_f=6q$ determinantal 't Hooft \cite{Hooft:1976} terms, for the u,d,s quarks, has no stable ground state. Global stability is implemented \cite{Osipov2:2006} by the addition of chiral invariant and OZI violating eight quark interactions. The multiquark picture is present in several approaches to low energy hadron dynamics: i) The instanton vacuum provides evidence in favour of $2N_f$-quark interactions (in the zero mode approximation). In leading $1/N_c$ order they are given by the 't Hooft determinant \cite{Hooft:1976}, which breaks the axial $U_A(1)$ symmetry and is a source of OZI-violating effects. ii) Non zero modes play an important role as well \cite{Simonov:1997} , to comply with the Banks and Casher result for chiral symmetry breaking \cite{Banks:1980}.
The effective quark Lagrangian derived from the instanton gas model, considered
beyond the zero mode approximation, predicts $4q, 6q, \ldots , 2nq, \ldots $ interactions, all equally weighted at large $N_c$ \cite{Simonov:2002}. iii) Lattice results for gluon correlators \cite{Bali:2001} reveal a hierarchy with dominance of the lowest ones; they could trigger a similar hierarchy in terms of the multiquark interactions, after integrating out the gluonic degrees of freedom.
Within our model the hierarchy in multiquark interactions is made possible in the large $N_c$ counting scheme. The minimal constellations which support the symmetry principles of low energy QCD and have a globally stable vacuum involve $4q$, $6q$ and $8q$ interactions. We suppose that these interactions are localized in the interval $\Lambda_{conf}<\Lambda
<\Lambda_{\chi SB}$ of confining and chiral symmetry breaking scales. The effects of multiquark terms on the vacuum are easily illustrated by calculating the scalar effective potential in the  $SU(3)$ chiral limit (the general case $m_u\ne m_d\ne m_s$ is derived in \cite{Osipov2:2006}). Results are displayed in fig. 1, as function of the order parameter for chiral symmetry breaking. The upper panel represents the Wigner-Weyl, the lower panel the phase of spontaneous chiral symmetry breaking, the corresponding curvature of the potential at the origin is not altered by inclusion of higher multiquark interactions, see details in caption and \cite{Osipov3:2006}. Spontaneous breakdown of chiral symmetry can also occur superimposed to the Wigner-Weyl phase, induced by the 't Hooft interactions (figure at far right).


\hspace{1.9cm} \framebox{$4q$}
\hspace{2.cm} \framebox{$4q + 6q$}
\hspace{2.2cm} \framebox{$4q+6q+8q$}

\vspace*{0.7cm}

\begin{figure}[h]
 \hspace{-3.5cm}  \includegraphics[height=.22\textheight]{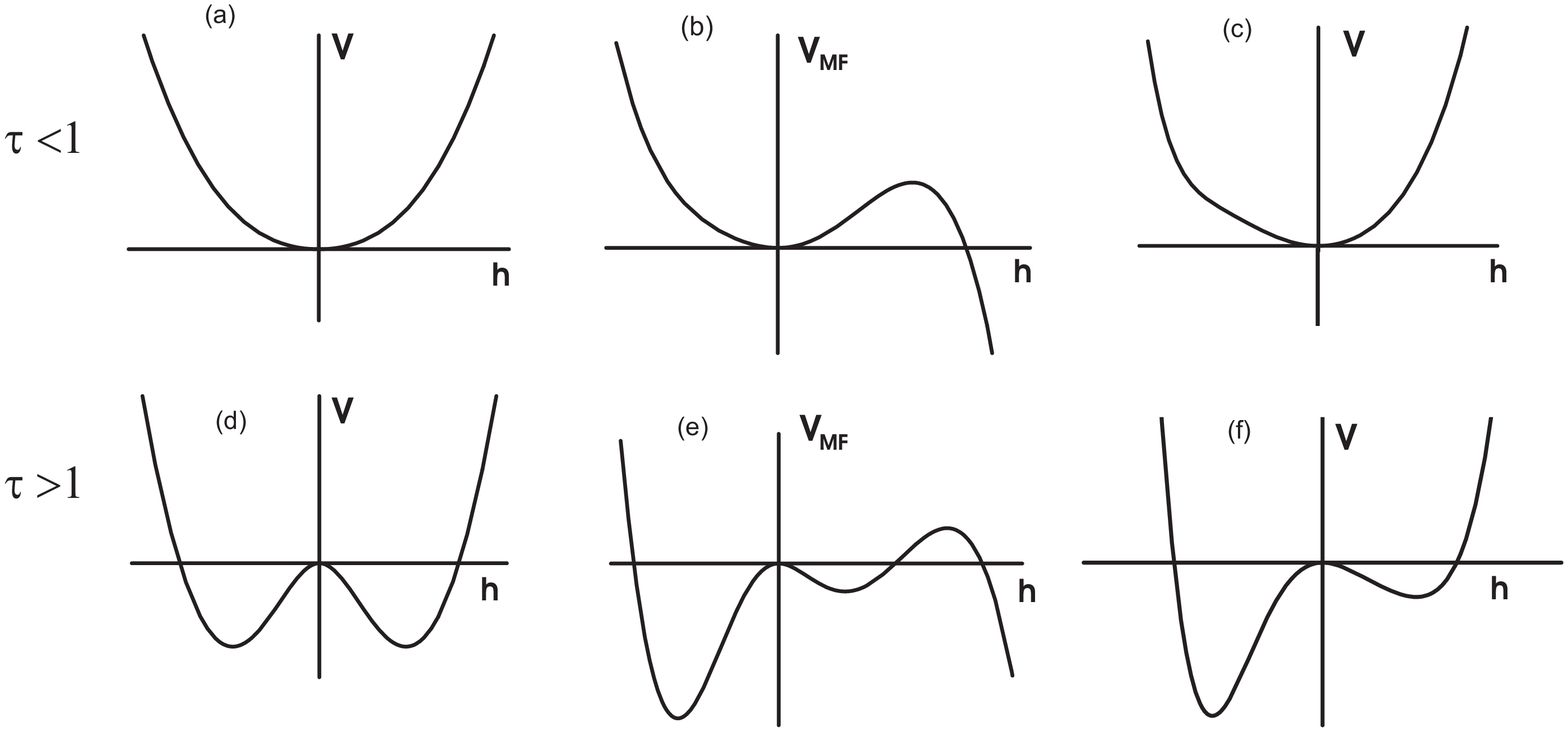}
  \caption{Ef\mbox{}fective potential V in the $SU(3)$ limit, calculated in the stationary phase approximation (SPA), $h \propto$  quark condensate, $\tau \propto$
curvature of V at origin. Each panel shows the typical form of the potential when one adds successively to the $4q$ case (a,d) the $6q$ (b,e) and $8q$ (c,f) terms. In (b,d) it is shown that the 't Hooft interaction ${\sim 6q}$ renders the NJL vacuum of (a,d) metastable in the mean field approximation, $V_{MF}$, (SPA leads to a vacuum without any local minimum; $V_{MF}=V$ only for stable configurations). Global stabilization is achieved by adding the $8q$ terms, figs. (c,f). Upper far right:a closer view of fig. 1c. A further mechanism of chiral symmetry breaking is at play at some critical value of $\kappa$. This phase coexists with the Wigner-Weyl phase.}
\end{figure}

\vspace*{-9.2cm}
\begin{figure}[h]
 \hspace{10cm} \includegraphics[height=.12\textheight]{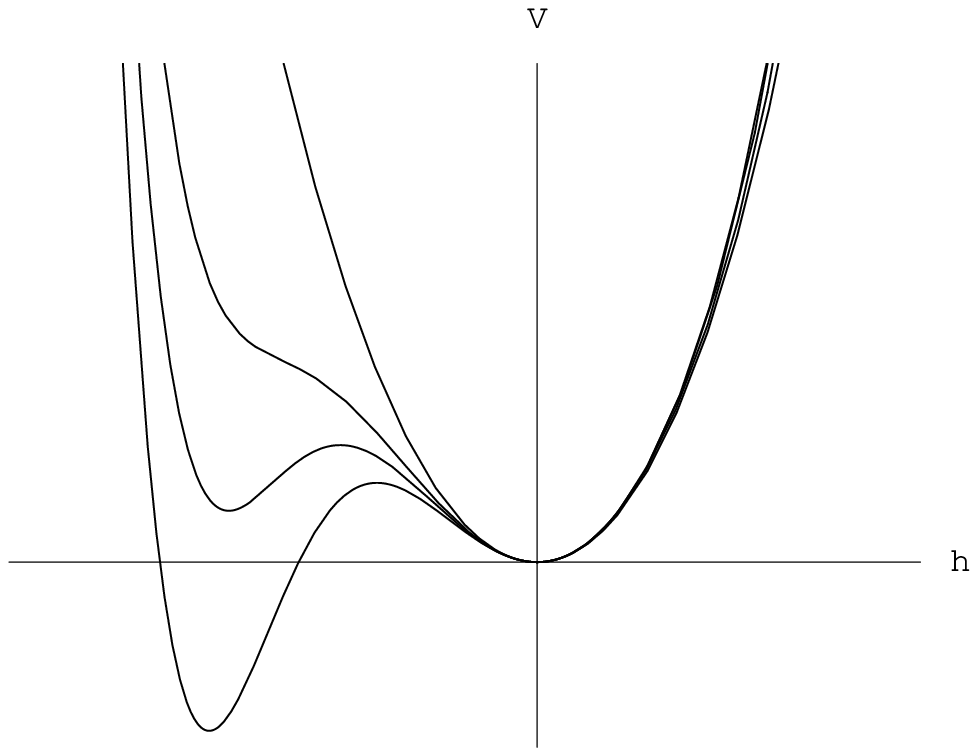}
\end{figure}
\vspace{6.3cm}

The $N_c$ counting rules for the different couplings of the model are $G\sim 1/N_c$, $\kappa\sim 1/N_c^3$,  $1/N_c^5 \leq g_1 \leq 1/N_c^4$, related with the $4q, 6q$ and $8q$ interactions respectively.
\vspace{0.3cm}

We now address the effects of the $8q$ forces on the low lying spin zero meson mass spectra. Here we discuss the trends, for numerical results see \cite{Osipov3:2006}. The effect on pseudoscalars  is small and observed in the $\eta-\eta'$ mass splitting,
\begin{equation}
\label{Venext}
   m_\eta^2 = m_0^2 -\frac{8(m_K^2-m_\pi^2)^2+
   3c_q }{9(m_{\eta'}^2-m_0^2)}\, ,
\end{equation}
\noindent
where $m_0^2 =\frac{1}{3}(4m_K^2-m_\pi^2)$ is the Gell-Mann -- Okubo result
for the $\eta$-mass. Numerically $m_0=565\,\mbox{MeV}$ is just a
bit larger than the phenomenological value
$m_{\eta} = 547.30\pm 0.12\, \mbox{MeV}$.
The remainder originates in the repulsion of
$\eta$ and $\eta'$ and is a $SU(3)$ breaking effect of second order.
The Witten-Veneziano correction, i.e., the second term
$\sim (m_K^2-m_\pi^2)^2$, is related to the topological susceptibility
and is about four times larger than required \cite{Witten:1979}, leading to a too low value for the $\eta$ mass.  The coef\mbox{}ficient  $c_q$  extends the
Veneziano result by including corrections from 't Hooft and $8q$
interactions \cite{Osipov3:2006}. They yield an additional small
correction $\sim 2\%$ to the Witten-Veneziano term.
In the large $N_c$ limit the $\eta,\eta'$ masses coincide with the result of Veneziano and Witten and $8q$ effects are absent in the leading order result for the topological susceptibility, although formally they could contribute \cite{Osipov3:2006},
\begin{equation}
\label{tops}
  \chi (0)|_{YM}  = \frac{\kappa}{4}
   \left(\frac{M}{2G}\right)^3
   \qquad (\mbox{large}\ N_c),
\end{equation}
\noindent where $M$ stands for the constituint quark mass in the $SU(3)$ limit.

In the scalar sector we get the following hierarchy within the nonet ($f_0^-,f_0^+$ stand for the singlet-octet mixed states) :
\vspace{0.5cm}

\centerline{$m_{f_0^-}<m_{a_0}<m_{K_0^*}<m_{f_0^+}$}
\vspace{0.5cm}

\noindent
which can be understood already at leading order of $N_c$, \cite{Osipov3:2006}.
So within the model restrictions (no confinement, no renormalizability) the scalar mesons built from one-loop quark-antiquark interactions do not conform with the empirical ordering of their masses. If the "ab initio" calculated $l_i$ of the model reproduce the values obtained in a recent investigation within unitarized CHPT, in which an imposed large $N_c$ limit helps to explore the quark-antiquark versus a more intricate meson structure \cite{Pelaez:2006}, the present Lagrangian yields further evidence in favour of a more complex structure.

A significant effect of $8q$ interactions is present in the low lying $\sigma$ meson ($f_0^-$), seen from the approximate sum rule \cite{Dmitrasinovic:1996} in a large $N_c$ estimate
\begin{equation}
   m_{\eta'}^2+m_\eta^2-2m_K^2 + m_{f_0^+}^2 +
   m_{f_0^-}^2-2m_{K_0^*}^2 =  -6E_1^{LO}
   +{\cal O}\left(\frac{1}{N^2_c}\right).
\end{equation}
 The $E_1$ term on the RHS has its origin in an $8q$ interaction term stemming from the mass relations ${f_0^{-,+}}$, and contributes at $1/N_c$ order, if  $g_1\sim 1/N_c^4$.
 This term has a negative sign, decreasing the sum
 $m_{f_0^-}^2 + m_{f_0^+}^2$. With increasing
$g_1$ mainly the value of $m_{f_0^-}$ is lowered \cite{Osipov3:2006} and the octet-singlet splitting
grows in the scalar nonet.
This sum rule is a good illustration of the possible
impact of the eight-quark OZI violating forces on the scalar mesons.

Research supported by FCT, Unidade I\&D 535, POCI/FP/63930/2005 and the EU integrated
infrastructure initiative Hadron Physics project
No.RII3-CT-2004-506078. A.A. Osipov gratefully acknowledges
Funda\ca o Calouste Gulbenkian for f\mbox{}inancial support.



\bibliographystyle{aipprocl} 


\end{document}